\documentclass[pra,twocolumn,aps,showpacs]{revtex4-1}

\usepackage{hyperref}
\usepackage{graphicx}
\usepackage{amsmath}
\usepackage{amsfonts}
\usepackage{amssymb}
\usepackage{subfigure}
\usepackage[usenames,dvipsnames]{color}
\usepackage{bm}
\everymath{\displaystyle}
\newcommand{\bea}{\begin{eqnarray}}
\newcommand{\eea}{\end{eqnarray}}
\begin{document}
\title{Acoustic radiation from vortex-barrier interaction in 
       atomic Bose-Einstein condensate}

\author{K. Suthar}
\affiliation{Physical Research Laboratory,
             Ahmedabad - 380009, Gujarat,
             India} 
\author{Arko Roy}
\affiliation{Physical Research Laboratory,
             Ahmedabad - 380009, Gujarat,
             India}
\author{D. Angom}
\affiliation{Physical Research Laboratory,
             Ahmedabad - 380009, Gujarat,
             India}
 

\date{\today}


\begin{abstract}
We examine the dynamics of a vortex dipole in the Bose-Einstein condensates 
(BECs) of trapped dilute atomic gases at zero temperature in the presence of 
a Gaussian barrier potential. The density-anisotropy induced by the barrier 
enhances the acoustic radiation from the vortex dipole. This is due to the 
deviation of the condensate density from the equipotential curves and 
variation in the curvature of the vortex dipole trajectory. Due to the 
acoustic radiation, the vortex dipole dissipates energy and spirals towards 
the edge of the condensate. As a result, we observe an increase in the 
vortex-antivortex annihilation events. To examine the effect of the Gaussian 
barrier, we estimate the correction to the Thomas-Fermi condensate density 
using perturbation expansion method and the results are in very good 
agreement with the numerical results. 
\end{abstract}



\maketitle

\section{Introduction}

The dynamics of topological defects like vortex in nonlinear systems is the
key to understand important phenomena in chemical patterns, fluid dynamics, 
liquid crystals, superfluids, etc ~\cite{pismen_99,kobayashi_07,henn_09}. 
In scalar Bose-Einstein condensates 
(BECs)~\cite{anderson_95,davis_95,bradley_95}, vortices carry integral 
angular momentum and serve as the obvious signature of superfluidity of these
systems~\cite{feynman_55,onsager_49}. The various experimental techniques 
which have been employed to generate vortices in BECs include manipulating 
the interconversion between the internal spin states of an 
isotope~\cite{matthews_99}, stirring the BEC with a laser
beam~\cite{madison_00}, rotating the BEC~\cite{aboshaeer_01},
phase imprinting~\cite{andrelczyk_01,leanhardt_02,freilich_10} and merging
of BECs~\cite{scherer_07}. Vortex dipoles, consisting of vortex-antivortex
pairs, have also been experimentally realized in BECs by moving the 
condensate across a Gaussian obstacle potential~\cite{neely_10}.
In a vortex dipole, vortices of opposite circulation cancel each other's 
angular momentum and thus carry only linear momentum. This is the cause of 
several fascinating phenomena such as leap frogging, snake 
instability~\cite{brand_02}, etc. Another important dynamical
phenomenon is the vortex-antivortex annihilation, which is expected to occur
when vortex and antivortex approach each other. There is, however, a dearth of
experimental signature. The introduction of a Gaussian barrier, examined in
the present work, ensures the vortex-antivortex annihilation occurs by
modifying the trajectories through acoustic radiation by the (anti)vortex.

On the theoretical front, among other
important phenomena, creation and dynamics of a vortex dipole in a BEC 
at zero temperature~\cite{pitaevskii_03,fetter_01,aioi_11}, lack of 
annihilation of vortex dipoles~\cite{prabhakar_13,gautam_14}, effect of an 
oscillating Gaussian potential~\cite{fujimoto_11}, and impact of the 
density inhomogeneity on the vortex motion~\cite{mason_08} have
been examined in previous works. The 
stability and dynamics of the clusters of vortices and antivortices in 
pancake-shaped BECs has also been studied~\cite{pietila_06}. It may also be 
mentioned here that in phase-separated binary condensates, coreless vortex 
dipoles can be formed by passing an obstacle across the condensate 
\cite{gautam_12} or changing the nonlinearities associated with the 
system~\cite{gautam_13}. The dynamics of a vortex dipole across an 
interface of quasi-2D two-component BEC has also been examined in 
Ref.~\cite{aioi_12}.

 Despite these significant experimental and theoretical advances, the
interactions between a single vortex or a vortex dipole with a barrier
(which can be experimentally achieved through a laser beam), within a BEC, 
and the associated vortex dynamics, are not fully understood and many 
problems remain unexplored. At zero temperature, in the absence of any 
other dissipative 
mechanism, the acceleration of a vortex generates acoustic radiation, which 
is the only sink of the turbulent kinetic energy~\cite{vinen_01,lundh_00}.
The decay of a single vortex, due to inhomogeneity induced acoustic
emission in a combined harmonic and Gaussian dimple trap potential has
been studied in a previous work~\cite{parker_04}. Similar studies have also
examined the acoustic radiation from multiple vortices of same circulation
or charge in a harmonic~\cite{barenghi_05} and double-well~\cite{parker_12}
trapping potentials.

 In the present work, we investigate theoretically the interaction of a
vortex dipole in a harmonically trapped quasi-2D BEC with a 
repulsive Gaussian barrier using the Gross-Pitaevskii
equation (GPE). The presence of the barrier breaks the rotational symmetry of 
the system and introduces novel effects on the dynamics of the vortex dipole. 
As a result of broken rotational symmetry, the equipotential and the 
isodensity curves do not coincide, hence the vortex or antivortex which 
precess along the equipotential curves traverse regions of different 
densities. When the vortex dipole encounters the barrier region, it emits 
acoustic radiation and some part of the kinetic energy is dissipated. 
This leads to a change in the trajectories of the 
vortex-antivortex pair which gradually spirals out of the condensate. The 
study of vortex dipole interaction with a barrier in a BEC will shed light 
on the process of dissipation of kinetic energy into acoustic energy, and 
as well as the conditions for annihilation along with other phenomena arising 
from the dynamics of vortex dipoles.

 The paper is organized as follows. In Section~\ref{vd_motion} we describe
the dynamics of vortex dipole using matched asymptotic expansion technique. 
In the same section we also discuss the correction to 
Thomas-Fermi(TF)-approximation near 
the Gaussian barrier, and examine the curvature and tidal effects. The
Section~\ref{numerical_result} is concerned with the numerical study of the 
dynamics of vortex dipole in the presence of the barrier in quasi-2D trap and 
the power radiated near the barrier. We, then, end with conclusions in 
Section~\ref{conclusions}.


\section{BEC with vortices}
\label{vd_motion}
In the mean-field approximation, the time dependent 
GPE describes the dynamics of a weakly interacting  BEC of dilute atomic 
gases very well. This implies $an^{1/3}\ll 1$ where, $a$ and $n$ are the 
$s$-wave scattering length of the atoms and density of the BEC, respectively. 
Thus, the order parameter of a single species scalar condensate, $\Psi$, is 
given by the GPE
\begin{equation}
  i\hbar\frac{\partial}{\partial t}\Psi(\mathbf r,t)
  = \left [-\frac{\hbar^2}{2m} \nabla^2 + V_{\rm ext}(\mathbf r) + 
     U|\Psi(\mathbf r,t)|^2
     \right ]\Psi(\mathbf r,t), 
\label{GP}
\end{equation}
where $V_{\rm  ext}$ is an external trapping potential, 
and $U = 4\pi\hbar^2 a/m$ is 
the strength of inter-atomic interaction with $m$ as the atomic mass. The 
order parameter $\Psi$ is normalized to the total number of atoms $N$ in the 
condensate. For the present study, $V_{\rm ext}$ consists of a harmonic 
confining potential $V_{\rm tr}$ and a repulsive Gaussian barrier potential 
along the $x$-axis $V_{\rm bar}$, which passes through the origin,
\begin{eqnarray}
 V_{\rm ext}(\mathbf r) & = & V_{\rm tr} + V_{\rm bar},\nonumber\\
                        & = &  \frac{m\omega_{\perp}^2}{2}(x^2 + y^2 
       + \alpha^2 z^2) + V_0\exp{\left ( -2\frac{y^2}{w^2}\right ) },
\end{eqnarray}
where $\omega_{\perp}$ is the radial trapping frequency, $\alpha$ is the 
anisotropy parameter ($\omega_z/\omega_\perp$), and  $V_0$ and $w$ are the 
amplitude and width of the Gaussian barrier, respectively. 
For the present work, $N$ is sufficiently large so that
$\mu\gg\hbar\omega_{\perp}, \hbar\omega_z$ and TF approximation is applicable
in the bulk of the condensate. In this approximation, the order parameter is
$\Psi_{\rm TF} = [(\mu - V_{\rm tr})/U]^{1/2}$. However, this approximation 
does not provide a good description in the regions where the condensate 
density has large gradients, and the kinetic energy is not negligible. We 
consider quasi-2D geometry (pancake shaped) of the trapping 
potential, which implies $\alpha \gg 1$ or $\omega_z \gg \omega_\perp$.


\subsection{Dynamics of vortices in BEC}

The hydrodynamic equations corresponding to the GPE represents an irrotational
flow. Vortices, however, may occur when there are phase singularities in 
$\Psi$. The motion of a vortex line in the condensate depends on the trapping 
potential, angular velocity of the trap and distortion of the vortex line. 
The expression for the vortex velocity has been calculated using 
coordinate transformations and the method of matched asymptotic expansion
~\cite{rubinstein_94,svidzinsky_00,nayfeh_08}, 
where the TF approximation is used in the matching region.
A vortex element carrying charge $q$ located at the position
$\mathbf{r}_0$ from the center of a rotating trap with angular 
velocity $\mathbf{\Omega}$, has velocity~\cite{koens_12} 
\begin{eqnarray}
 \mathbf v(\mathbf r_0) = &&\frac{\hbar q (\hat z\times\hat t)
        (\partial V_{\rm tr}/\partial z)|_{\mathbf r = \mathbf r_0}}
        {2mU|\Psi_{\rm TF}|^2}[\ln(\xi)+
        \hat{\rho}\otimes\hat{\rho}]\nonumber\\
    &+& \frac{\hbar q \kappa \hat b}{2m}\biggr[\ln\biggr(\frac{1}{\xi}
        \sqrt{\frac{-\rho_{\rm 0}}
         {\kappa \hat b \cdot\hat{\phi}}}\biggr)
        +\hat{\rho}\otimes\hat{\rho}\biggr]\nonumber\\
    &+& \frac{\hbar q[{\hat z}\times\nabla 
        V_{\rm tr}(\mathbf r_0)]}{2mU|\Psi_{\rm TF}|^2} \biggr[\frac{3}{2}
        \ln\biggr(\frac{R_{\perp}}{\xi}\biggr)+
        \hat{\rho}\otimes\hat{\rho}\biggr]\nonumber\\
    &+& 2\frac{\nabla V_{\rm tr}(\mathbf r_0)\times\mathbf \Omega}
        {\nabla_\perp^2V_{\rm tr}(\mathbf r_0)}, 
 \label{vel_eqn}
\end{eqnarray}
where $\hat t$, $\hat n$ and $\hat b$ are the 
tangent, normal and binormal vectors, collectively form the Frenet-Serret 
coordinate system. These are orthogonal vectors which form a basis of the 
local coordinate system centered at the vortex element. 
Any point on the vortex line at a radial distance $\rho_0$ 
from the trap center is represented by ($\rho$, $\phi$, $z$) in cylindrical
coordinates. The unit vectors $\hat{\rho}\approx\mathbf{r}/\rho$ and 
$\hat{\phi}$ are along radial and azimuthal directions, defined for the 
far-field region with respect to the vortex element. The parameter 
$R_{\perp}$ is the radial extent of $\Psi_{\rm TF}$, $\kappa$ is the 
curvature of the vortex line, and $\xi = \hbar/\sqrt{2mnU}$ is the healing 
length, where $n=|\Psi_{\rm TF}|^2$ is the local density of the condensate.
The parameter $\xi$ is an important quantity as it is a measure of the vortex 
core size. Here, the symbol $\otimes$ represents the tensor product 
of two radial unit vectors. 

 In the present work, we consider a quasi-2D nonrotating trapped condensate 
for which $\mathbf{\Omega} = 0$. We can also assume $\kappa\approx$ 0, since 
the curvature of the vortex lines do not play any significant role in a 
quasi-2D system.  Thus the dynamics of a vortex dipole in quasi-2D condensate 
is solely governed by the third term of  Eq.~(\ref{vel_eqn}). 
Based on which, a single off-axis vortex moves along an 
equipotential curve and the vortex 
precess around the trap center. In the case of multiple vortices in the 
condensate, the inter-vortex induced velocity must also be included. 
Here, we consider the case of a vortex-antivortex pair or a vortex dipole, 
with the vortex and antivortex located at $\mathbf{r}_1$ and $\mathbf{r}_2$, 
respectively. Thus, the velocity component of the vortex due to the presence 
of the antivortex is given by
\begin {eqnarray}
 \mathbf v_{\rm {12}} & = & -q\frac{\hbar}{m}
                              \nabla \phi_2(\mathbf{r}_1),\nonumber\\
                      & = &  q^2\frac{\hbar}{m}
                             \frac{[\nabla\times(|\Psi_{\rm TF}|^2
                             \Phi(\mathbf{r}_1,\mathbf{r}_2){\hat{z}_1})]}
                             {|\Psi_{\rm TF}|^2},
 \label{ind_vel}
\end {eqnarray}
where $\phi_2$ is the phase of the condensate associated with the antivortex. 
In above equation we can write $\nabla \phi_2$ at the location of vortex as 
$-q[\nabla\times(|\Psi_{\rm TF}|^2\Phi(\mathbf{r_1},
\mathbf{r_2}){\hat{z}_1})]/|\Psi_{\rm TF}|^2$, with $\hat{z}_1$ being 
the unit vector in the vortex's frame. This form is chosen such that it 
satisfies the property $\nabla\cdot(|\Psi_{\rm TF}|^2\nabla \phi_2)=0$
~\cite{koens_12}. 
The {\em pseudo-vector potential} in the above equation is defined as
\begin{equation}
 \Phi(\mathbf{r}_1,\mathbf{r}_2)   
                   = -K_0 \Bigg(\sqrt\frac{\nabla_{\perp}^2 V_{\rm tr}}
                     {2U|\Psi_{\rm TF}|^2}|\mathbf{r}_1 -\mathbf{r}_2|\Bigg),
\end{equation}
where $K_0$ is the modified Bessel function of the second kind. 

 During the dynamical evolution, the vortex and antivortex are closest when 
they traverse the bulk of the condensate where TF-approximation is 
applicable. Within the bulk of the condensate, 
$\nabla_{\perp}^2 V_{\rm tr}/(2U|\Psi_{\rm TF}|^2)\approx 1/R_{\perp}^2$.
For small separation between vortex and antivortex,
\begin{equation}
 \Phi(\mathbf{r}_1,\mathbf{r}_2)\approx\ln\biggr(\frac{e^{\gamma}|
      \mathbf{r}_1 - \mathbf{r}_2|}{2R_{\perp}}\biggr),
\end{equation}
where we have used $K_0(x)\approx -\ln\left(e^{\gamma}x/2\right)$ for 
$x\rightarrow 0$~\cite{wolf_10}and $\gamma$ is the Euler constant. 

 The trajectory of a vortex dipole results from the competition between the 
effects of vortex-antivortex interaction and precession~\cite{li_08}. 
The net velocity of the vortex or antivortex is the sum of
the individual velocity, third term of Eq.~(\ref{vel_eqn}), and the mutual 
velocity field given by Eq.~(\ref{ind_vel}).  In the calculation of the 
individual velocity, we neglect $\hat{\rho}\otimes\hat{\rho}$ as it is 
smaller than the logarithmic term~\cite{koens_12}. 
Additionally, the contribution from mutual velocity field is
dominant when the vortex-antivortex pair is closely separated. This happens
in the bulk-region of the condensate where $|\Psi_{\rm TF}|^2$ is assumed
to be constant. Thus, the dynamics of a vortex dipole consisting of a singly 
charged vortex ($q=1$) and antivortex ($q=-1$) located at ($x_1,y_1$) and 
($x_2,y_2$), respectively, is given by the following coupled differential 
equations.
\begin{eqnarray}
 \mathbf v_i & = &  (-1)^{i+1} \frac{3\hbar
                   [{\hat z_i}\times\nabla_i V_{\rm tr}(\mathbf r_i)]}
                    {4mU|\Psi_{\rm TF}|^2} 
                    \ln\bigg(\frac{R_{\perp}}{\xi}\bigg)\nonumber\\
              &&    + \frac{\hbar}{m} \bigg[\nabla_i \times
                      \ln\biggr(\frac{e^{\gamma}|\mathbf r_i
                    - \mathbf r_j|}{2R_{\perp}}\biggr)
                      \hat z_i\bigg],
 \label{vel_dip1}
\end{eqnarray}
where $i=$ 1 and 2 represents the vortex and antivortex, respectively, 
and $j = 3-i$. After simplification, we get
\begin{eqnarray}
 \mathbf v_i  & = & (-1)^i\frac{3\hbar\omega^2_\perp}{4U|\Psi_{\rm TF}|^2}
                     (y_i\hat x_i - x_i \hat y_i)
                     \ln\bigg(\frac{R_{\perp}}{\xi}\bigg)\nonumber\\
              &&    + \frac{\hbar}{m}
                      \bigg[\frac{(y_i - y_j)\hat x_i -
                      (x_i - x_j)\hat y_i}
                      {|\mathbf r_i - \mathbf r_j|^2}\bigg],
 \label{vel_dip2}
\end{eqnarray}
where the unit vectors $\hat x_i$, $\hat y_i$ and 
$\hat z_i$ are defined along the local coordinates of the vortex and 
antivortex. 

 Now, we consider the effect of the Gaussian barrier potential on the dynamics 
of the vortex dipole. Since we introduce 
the barrier along $x$-axis, it affects the $x$ component of the 
velocities when $|y_i|\leqslant w$, while the $y$ component of the velocities 
remain unaffected. This is because from Eq.~(\ref{vel_dip1}), using 
$V_{\rm ext}$ in place of $V_{\rm tr}$, the velocity of the vortex or 
antivortex depends on 
$(-1)^{i+1}\hat{z}_i\times\nabla V_{\rm ext}(\mathbf r_i)\approx 
(-1)^{i+1}\hat{z}_i\times(\partial V_{\rm bar}/\partial y_i)\hat{y}_i$. 
The vortices cross the barrier two times in an orbit, which is evident from 
the trajectories shown in  Fig. \ref{traj_vd}(b).  The first and second 
crossings occur when vortex-antivortex are far and closely separated, 
respectively. The effect of the barrier potential is more prominent in the 
former as the intervortex interaction is negligible. In this case within the 
barrier region ($|y_i|\leqslant w $) and neglecting intervortex interaction,
the velocities along the $x$-axis are
\begin{equation}
 \mathbf v_{ix} = (-1)^i\frac{3\hbar\omega^2_\perp}{4U|\Psi_{\rm TF}|^2} 
                    \bigg(1-\frac{4V_0}{m\omega^2_\perp w^2}\bigg)y_i
                      \ln\bigg(\frac{R_{\perp}}{\xi}\bigg) \hat{x}_i,
 \label{vel_x}
\end{equation}
where we have retained only the first term in the expansion of $V_{\rm bar}$.
Away from the barrier ($|y_i|>w$), the influence of the barrier potential
is weak and may be neglected. The modified velocities, thus, depend on the 
barrier parameters $V_0$ and $w$. We choose $V_0$ sufficiently 
large ($V_0\gg m\omega^2_\perp w^2$) such that it affects the dynamics of 
the vortices in the quasi-2D BEC and in particular, the velocity component 
along $x$-direction. 


\subsection{Correction to TF approximation}

The presence of the barrier along $x$-direction in the 
condensate introduces a large density gradient in $y$-direction. 
Thus, for $|y_i|\leqslant w $ the TF approach fails to 
describe the system and this motivate us to calculate the 
deviation from TF approximation. The leading order correction to the TF 
approximation may be calculated using standard perturbation expansion 
techniques~\cite{lundh_96,dalfovo_96,fetter_98}. In the present work we 
examine the correction to the TF density profile along $y$-direction due to 
the presence of the Gaussian barrier using an approach similar to ref. 
\cite{lundh_96}. 

 To describe the structure of $\Psi$ near the repulsive barrier, we consider 
the condensate in the quasi-2D limit ($\alpha \gg 1 $). In this regime, 
we approximate the equilibrium order parameter as 
$\Psi(x,y,z, t) = \exp(-i\mu t/\hbar) \psi(x,y)\phi_{0}(z)$, 
where $\phi_{0}(z) = [\alpha/2\pi]^{1/4}\exp\left(-\alpha z^2/4\right)$ is 
the ground state wave function along the $z$-direction and $\mu$ is the 
chemical potential. Using this ansatz in Eq.~(\ref{GP}) and integrating over 
$z$-direction, we obtain the dimensionless 2D GPE
\begin{eqnarray}
  & & \bigg[-\frac{1}{2}\bigg(\frac{\partial^2}{{\partial \tilde{x}}^2}+
      \frac{\partial^2}{{\partial \tilde{y}}^2}\bigg) 
      + \frac{1}{2}(\tilde{x}^2 + \tilde{y}^2) + 
    \tilde{V_0} \exp\bigg(-2\frac{\tilde{y}^2}{\tilde{w}^2}\bigg) \nonumber\\
  & & + \tilde{g} |\tilde\psi(\tilde x,\tilde y)|^2\bigg]
        \tilde\psi(\tilde x,\tilde y) 
      = \tilde\mu\tilde\psi(\tilde x,\tilde y),
  \label{gp_2d}
\end{eqnarray}
where we use the oscillator length 
$a_{\rm osc}=\sqrt{\hbar/(m\omega_\perp)}$, 
$\omega_{\perp}^{-1}$, and the oscillator energy $\hbar\omega_\perp$ as the 
dimensional units of length, time and energy, respectively. The 
effective dimensionless 2D interaction strength 
$\tilde{g} = 2 \sqrt{\pi\alpha} (a/a_{\rm osc})$. Here, $\tilde\mu$
is the shifted chemical potential in scaled units obtained after integrating 
out the axial direction $z$ from the GPE in 3D. For notational simplicity, we 
will represent the scaled quantities without tilde in the rest of the 
manuscript.

 Without the barrier potential the equipotential curves coincide with the 
isodensity curves and a single vortex (antivortex) precess along these curves 
in anticlockwise (clockwise) direction. As the barrier potential 
is along $x$-axis, and $V_0/w^2\gg 1$, for $|y|/w\ll 1$ we have 
$\partial V_{\rm ext}/\partial x \ll \partial V_{\rm ext}/\partial y$. So,
in this region we can consider $\nabla_\perp V_{\rm ext}\approx 
\hat{y} \partial V_{\rm ext}/\partial y $ with $\hat{y}$ 
defined as the unit vector along the $y$-direction and the vortices undergo 
a large change in velocity. The Eq.~(\ref{vel_x}) shows that, 
the velocity of the vortex and antivortex is dependent on the $y$-coordinate
within the barrier region. When the vortex or antivortex enters the barrier 
region, it experiences a sharp deceleration for $y\sim w$ but the
velocity drops to zero at $y=0$ and again for $y\sim -w$, it experiences
an acceleration till it leaves the barrier region. This is confirmed with 
numerical computations for the antivortex velocity.

 To estimate the leading order correction to $\psi_{\rm TF}$, we consider the 
case of $ V_0 = \mu$, so that $\psi_{\rm TF}$ is zero at the origin. This 
assumption simplifies the calculation of the correction in the 
condensate order parameter near the barrier.
Thus, for $|y|/w\ll1$ with the above considerations at a fixed value of $x$, 
Eq.~(\ref{gp_2d}) simplifies to 
\begin{eqnarray}
  & & \bigg[-\frac{1}{2}\frac{\partial^2}{{\partial y}^2}+\frac{1}{2}y^2 + 
     V_0 \exp\bigg(-2\frac{y^2}{w^2}\bigg)  \nonumber\\
  & & + 2a\sqrt{\pi\alpha}|\psi(y)|^2 \bigg]\psi(y)= \mu\psi(y),
 \label{gp_1d}
\end{eqnarray}
where we have redefined $\mu $ by subsuming the
$x$ component of the trapping potential.  We introduce a 
scaled length variable $\zeta = y/\delta$, where $\delta$ is 
the distance between the center of the trap and classical turning points
(where total energy is equal to the potential energy) near the barrier in 
$y$-direction. In the $|y| < \delta$ region the kinetic energy is not 
negligible and this is the prime reason for calculating the correction to 
TF-approximation. In this regard, $\delta$ is a characteristic length which 
is associated with the structure of the barrier and number of atoms in the 
system. The scaled variable is such that $\delta\ll 1$ and $\zeta\gg 1$, but 
$\delta\zeta\ll w$. In this domain, the barrier potential 
$V_0 \exp(-2y^2/w^2)\approx V_0 (1 - 2\delta^2\zeta^2/w^2)$ and 
Eq.~(\ref{gp_1d}) becomes 
\begin{equation}
 -\psi'' - \mu'\psi+\eta\zeta^2\psi+\nu\psi^3 = 0,
 \label{scaled GP}
\end{equation}
where the prime denotes the differentiation with respect to the scaled 
variable $\zeta$, $ \mu' = 2(\mu- V_0)\delta^2$, 
$\eta = (1-4 V_0/w^2)\delta^4$ and $\nu = 4 a\delta^2\sqrt{\pi\alpha}$.
Thus the solution with the TF-approximation is
\begin{equation}
 \psi_{\rm TF} = \bigg(\frac{\mu'-\eta\zeta^2}{\nu}\bigg)^{1/2}  \text{ for }
        \zeta\gg1.
 \label{TF_sol}
\end{equation}
To determine the leading order correction to $\psi_{\rm TF}$, we define
$\psi = \psi_{\rm TF} + \psi_1$ and treat $\psi_1$ as a correction 
arising from the presence of the barrier along $x$-direction. After linearizing 
Eq.~(\ref{scaled GP}) and only considering terms linear in $\psi_1$, we get
\begin{equation}
 -\psi_1''-(\mu'-\eta\zeta^2)\psi_1+3\nu\psi_{\rm TF}^2\psi_1 = \psi_{\rm TF}''.
\end{equation}
Now, after neglecting $\psi_1''$, as it contributes to terms of higher order 
in $1/\zeta^2$, the correction $\psi_1$ is 
\begin{equation}
 \psi_1 = \frac{-\eta\mu'}{2\nu^{1/2}}\frac{1}{(\mu'-\eta\zeta^2)^{5/2}}.
 \label{TF_corr}
\end{equation}
Thus, the corrected solution of the GPE in Eq.~(\ref{gp_1d}) with the 
TF-approximation is 
\begin{equation}
 \psi = \bigg(\frac{\mu'-\eta\zeta^2}{\nu}\bigg)^{1/2}\left[1-\frac{\eta\mu'}
        {2(\mu'-\eta\zeta^2)^3}\right].
\end{equation}
With this improved solution, the equipotential and isodensity curves do not 
coincide anymore and as a consequence, the vortex or antivortex which precess 
along the equipotential curves traverse regions of different densities. 
Due to the density inhomogeneity, the vortex or the antivortex experiences a
change in the velocity and generates acoustic radiation. In other words,
there is a change in the kinetic energy of the vortex or antivortex as a part 
of it is transformed into acoustic radiation energy.


\subsection{Curvature and tidal effects}

In addition to the change in velocity arising from the density gradient, 
the vortex or antivortex also undergoes acceleration due to the curvature of 
the equipotential curves of $V_{\rm ext}$. With only the harmonic oscillator 
potential, the equipotential curves are circles and have constant curvature. 
However, in the presence of the barrier potential, the equation of 
equipotential curve with energy $C$ in 2D is 
\begin{equation}
 \frac{1}{2}(x^2+y^2)+ V_0 \exp\left(-2\frac{y^2}{w^2}\right) = C, 
 \label{equipot_curve}
\end{equation}
where $C\leqslant R^2_\perp/2$. The equation of the 
curve can be rewritten as a function of $y$ and we can then calculate the 
curvature at a point on the curve. 
From basic differential geometry, for any two points with the same
abscissa $x$, and $\pm y$ as the $y$-coordinate on an equipotential curve, 
the curvature is given by 
\begin{eqnarray}
 K = \bigg|\frac{d\varphi}{ds}\bigg|
   = \left [2C-2 V_0 \exp\left(-2\frac{y^2}{w^2}\right)\right]^{-1/2},
 \label{curvature}
\end{eqnarray}
where $\varphi$ and $s$ are the angle subtended by the tangent to the 
$x$-axis and arc length of the curve. This indicates that for $|y|\ll w$  
we get $K \approx (2C + 4V_0y^2/w^2 - 2V_0 )^{-1/2} $, which can be large 
when $|y|/w\sim 1$. So, when the vortex or antivortex traverses this region 
it undergoes large acceleration and induces acoustic radiation. In the 
region where $|y|> w$ the effect of barrier is negligible and 
$K \approx 1/\sqrt{2C}$. Due to the constant $K$, as mentioned earlier, 
the vortex or antivortex undergoes uniform acceleration and generates 
acoustic radiation~\cite{parker_04}. However, the power radiated is much 
smaller than in the region with large changes in $K$. Later, we shall 
examine the temporal variation of the radiated energy $J$ in detail with 
numerical simulations. 

 To illustrate the dissipation process, we consider the case of a single 
vortex passing through the barrier potential. The velocity of vortex near the 
barrier can be understood in terms of the vortex velocity field given in the 
Eq.~(\ref{vel_x}). According to the equivalence between a 2D superfluid and 
(2+1)D electrodynamic system, the vortices, superfluid density, and particle 
current play the roles of charges, magnetic field and electric field, 
respectively. Most importantly, the acoustic waves in the superfluid are 
analogous to the electromagnetic radiations~\cite{arovas_97}. The acoustic 
radiation from the vortex, due to large $K$, near the barrier is similar to 
the Larmor radiation from an accelerating charged particle~\cite{kambe_86}. 
The energy radiated in the far field is 
\begin{equation}
 J = \lim_{r\to\infty}\int_0^{2\pi} \mathbf{S}\cdot\hat{n} r\,d\theta,
 \label{en_flux}
\end{equation}
where the Poynting vector in the leading order is $\mathbf{S} = 
(\partial\phi/\partial t)\nabla\phi$~\cite{nore_97} with $\phi$ as the 
phase of the condensate, $r = \sqrt{x^2+y^2}$ 
is the circle enclosing the vortex and $\hat{n} $ is the normal to the circle 
at a point. We analyze the radiation in the regions far from the 
location of the vortex and antivortex. This is done by using far-field 
approximation, where the radiation emitted are similar to the acoustic
waves~\cite{krstulovic_08}. In the far-field limit 
$\nabla\phi\cdot\hat{n} \approx \dot{\phi}/c$, where 
$c= \sqrt{|\psi|^2g}$ is 
the acoustic speed and thus, the power radiated depends on the temporal
variation of $\phi$. To examine the temporal variation of $\phi$, 
we resort to the hydrodynamic description of the condensate. This is 
equivalent to the inviscid hydrodynamic equations with an additional term to 
represent the quantum pressure. The latter, quantum pressure, is important at 
the length scales of $\xi$ or smaller and  for the system of current interest
$\xi\sim 0.2-0.4\mu m$. This range of $\xi$ is derived 
from the density in the bulk region through the relation 
$\xi = 1/\sqrt{2 |\psi|^2 g}$, where we use $\psi$ obtained from the 
numerical solution of the GPE. 

 Since the barrier potential is along $x$-axis, we can neglect the density 
variation along $x$-axis in comparison to the $y$-direction. For a fixed 
value of $x$ we can write $\psi(y,t)|_x = \sqrt{n(y,t)/n}\exp[i\phi(y,t)]$, 
with $n$ as the average condensate density. The
dimensionless GPE can then be recast as two coupled equations of $n$ 
(in $y$-direction) and $\phi$ as,
\begin{subequations}
 \begin{eqnarray}
  {\dot n}   &=& -\partial_y \left[ n 
                     \partial_{y} \phi\right], \\
  {\dot\phi} &=& -\bigg [ V_{\rm ext} + n g - 
                    \frac{1}{2} \frac{\partial_y^2 \sqrt{n}}{\sqrt n}
                      \nonumber\\
               && + \frac{1}{2}\left(\partial_y \phi\right)^2 \bigg].
 \end{eqnarray}
\end{subequations}
In the second equation, the term with $\sqrt{n} $ arises from the quantum 
pressure. The other effect of the barrier arises from the large 
$\partial V_{\rm ext}/\partial y$ when $|y|\leqslant w $. Due to this large
gradient, two points within the vortex core separated by $\Delta y \ll 1$ 
have large difference in velocity. This distorts the geometry of the vortex 
core from circle to ellipse as it enters the domain of the barrier potential. 
The velocity difference between two points of the vortex 
separated by $2\xi$ along $y$-axis, based on the first term in 
Eq. (\ref{vel_dip1}), is
\begin{eqnarray*}
 \Delta v &=& v(x, y-\xi) - v(x, y+\xi)\\ 
          &=& \frac{3}{2}{\xi^2}\bigg[\hat{z}
              \times\bigg(\nabla V_{\rm ext}(x,y-\xi)\\
          &&  -\nabla V_{\rm ext}(x,y+\xi)
              \bigg)\bigg]\ln\bigg(\frac{R_{\perp}}{\xi}\bigg),
\end{eqnarray*}
and finally using the scaled expression for the $V_{\rm ext}$, we get
\begin{eqnarray}
 \Delta v &=& \frac{3\xi^3}{w^4}\bigg [\exp\left (\frac{2y^2+2\xi^2}
              {w^2}\right)(w^4-16y^2\xi^2) -4(w^2 \nonumber \\
          && -4y^2)V_0\bigg ]
           \exp\left(\frac{-2y^2-2\xi^2}{w^2}\right)
            \ln\bigg(\frac{R_{\perp}}{\xi}\bigg).
\end{eqnarray}
This finite difference in velocity across the vortex core near the barrier
gives rise to tidal effect. Furthermore, the core also acquires higher 
multipole moments due to the deviation from circular geometry. The shape of 
the core is well described by the ratio $n_v/n_0$, where $n_v$ and 
$n_0$ denotes the density of the condensate with and without vortex, 
respectively~\cite{jezek_08a}. It must be mentioned that, the vortex precession 
velocity depends on the background velocity field evaluated around the core 
and the shape of the vortex core~\cite{jezek_08b}. So, the barrier height and 
distortion of the core due to the tidal effect have direct bearing on the 
vortex dynamics. We shall examine it in more detail with the numerical 
computations.


\section{Numerical results}\label{numerical_result}
In the numerical simulation, we solve the dimensionless GPE using the 
Crank-Nicholson method~\cite{muruganandam_09} with following set of 
parameters: we consider $^{87}$Rb atoms with $a = 99 a_0$ (where $a_0$ is the 
Bohr radius), $N$ = $1\times 10^5$, $\alpha=25$ and 
$\omega_\perp = 2\pi\times10$Hz. We use phase numerical implementation of the
imprinting method~\cite{dobrek_99} to generate vortex dipole in BEC. For this, 
we begin the simulation with imaginary time propagation of the GPE and imprint 
the phase corresponding to a vortex dipole by using
\begin{eqnarray}
 \psi(x,y) & = &|\psi(x,y)|
               \exp\left[iq\tan^{-1}\left ( \frac{y-y_0}{x-x_0}\right) 
                \right] \nonumber \\
           && \times \exp\left[-iq\tan^{-1}\left ( \frac{y-y_0}{x+x_0}\right)
             \right ],
 \label{phase_imprint}   
\end{eqnarray}
where $(\pm x_0, y_0)$ are the location of the vortex and antivortex of the 
vortex dipole, and $q$ is the charge of the imprinted vortices. Since singly 
charged vortices are energetically favorable compared to the multiply 
charged vortices~\cite{pethick_08}, we shall consider only singly charged 
vortices ($|q|$=1).

\begin{figure}[h]
 \centering
 \includegraphics[width=9.0cm]{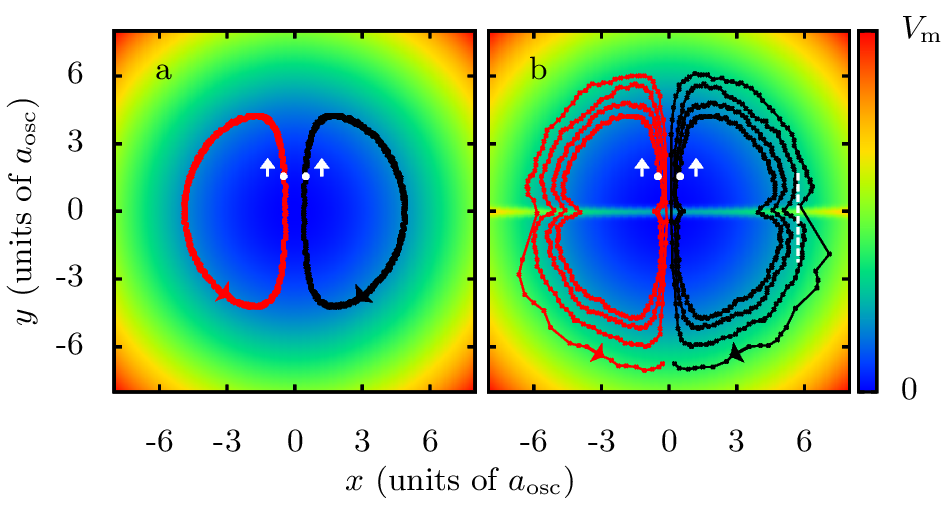}
 \caption{(Color online) Motion of a vortex dipole in an external trapping
          potential. (a) shows the trajectory of the vortex dipole in 
          the absence of the barrier potential, (b) shows the 
          trajectory of the vortex dipole in the presence of the barrier 
          potential. The initial position ($\mp 0.5$, $1.5$) 
          and direction of the motion of the vortex and antivortex are shown 
          by white dots and arrows, respectively. The initial separation of 
          vortex and antivortex is $1.0~a_{\rm osc}$. The red (black) lines 
          show the trajectories of vortex (antivortex) where direction of 
          the movement of the vortices is shown by the arrows.
          The straight dashed white line in (b) shows the decrease in
          the curvature of equipotential lines when the antivortex crosses 
          the barrier potential. The maxima of external trapping potential
          $V_{\rm m}$ is 128 $\hbar\omega_{\perp}$.}
 \label{traj_vd}
\end{figure}
\begin{figure}[h]
 \centering
 \includegraphics[width=9.0cm]{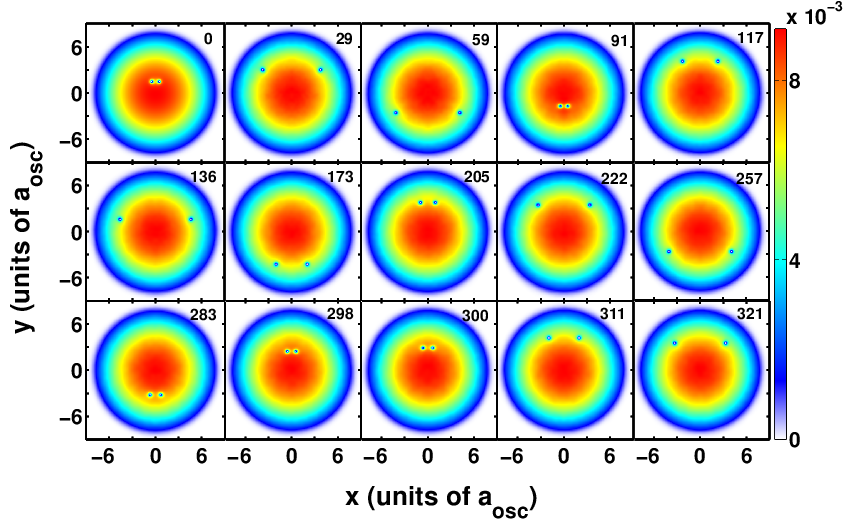}
 \caption{(Color online)  Condensate density profiles showing 
          the dynamics of a vortex dipole in a harmonically trapped BEC 
          (in the absence of the barrier potential) obtained using numerical 
          integration of 2D GPE. The vortex and antivortex are imprinted at 
          ($x_0 = \mp 0.5$, $y_0 = 1.5$). The vortex dipole completes its 
          first, second and third orbit at around 
          $96$ ms, $187$ ms, and $292$ ms respectively. The time (in units
          of ms) is shown at the top right corner of each image. 
          Density is measured in units of $a_{\rm osc}^{-2}$ and is
          normalized to unity.}
 \label{dipole_orbit}
\end{figure}

To study the dynamics, we consider the converged solution from the imaginary 
time propagation as the initial state and propagate it in real time. For the 
axisymmetric trap considered here, we observe the acceleration of vortices 
due to their mutual velocity fields which induces emission of acoustic waves. 
During the dynamical evolution, as the vortex-antivortex pair approaches the 
origin their separation decreases and appears to coalesce but do not 
annihilate~\cite{prabhakar_13}. After wards, they separate and continue to 
move. Thus, the vortex dipole exhibits periodic orbital motion and the 
position of the vortices at different times during three orbital cycles are 
shown in Fig. \ref{traj_vd}(a) and Fig. \ref{dipole_orbit}, respectively. We 
observe that the 
vortex dipole survives for many seconds without annihilation which was 
reported in our previous work~\cite{prabhakar_13} and is consistent with 
the experimental results~\cite{neely_10}. 

\subsection{Effect of barrier potential}
\begin{figure}
 \centering
 \includegraphics[width=8.5cm]{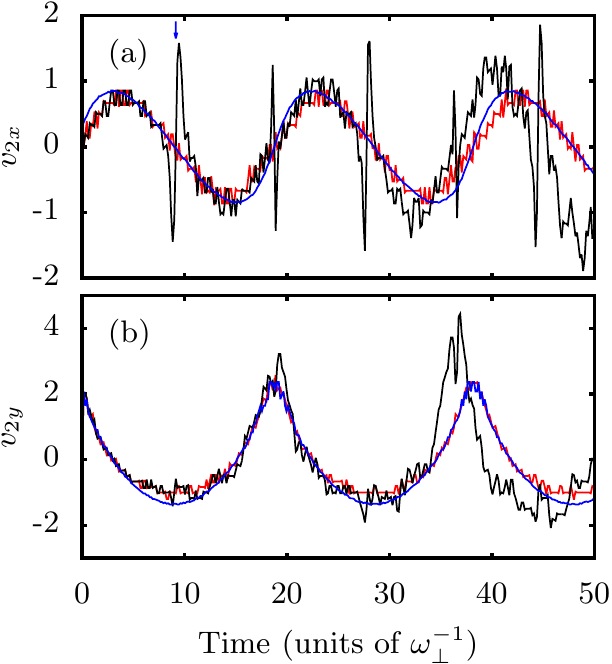}
 \caption{(Color online) The velocity components of the antivortex 
          with (without) the barrier potential. (a) Black (red) curve 
          represents the temporal variation of $v_{2x}$ 
          with (without) the barrier potential. The blue 
          arrow indicates the instant of time when $v_{2x}$ is zero.
          (b) Black (red) curve represents temporal variation in $v_{2y}$.
          Blue curve shown is the analytical result obtained from  
          Eq.~(\ref{vel_dip2}). Here the velocity components are expressed 
          in units of $a_{\rm osc}\omega_{\perp}$.} 
 \label{vel_comp}
\end{figure}

To examine the interaction of a vortex dipole with barrier, we introduce 
repulsive Gaussian barrier potential $V_{\rm bar} $ along the $x$-direction.
In the present study, we vary $V_0$ but set $w= 1.0 \mu$m. The trajectories 
of the vortex and antivortex in the presence of the barrier is shown 
in Fig. \ref{traj_vd}(b) for the case of $V_0=15$ (in units of 
$\hbar\omega_\perp$). 
Further more, the temporal variation of $v_{2x}$, the $x$-component  of the 
antivortex velocity, with and without the barrier are shown as black and red 
curves, respectively, in Fig. \ref{vel_comp}(a). As expected, without the 
barrier, $v_{2x}$ has periodic variation as the vortex precess in a 
semicircular trajectory and the first half-sinusoid in Fig. \ref{vel_comp}(a) 
represents $v_{2x}$ for the $y >0 $ part of the first orbit. Similarly, 
the second half-sinusoid correspond to the $v_{2x}$ for the $y<0 $ part. 
The zero of $v_{2x}$ occurs when the position of the antivortex has 
$y\sim 0$, which is consistent with the expression of $\mathbf{v}_2$ in
Eq.~(\ref{vel_dip2}).

The temporal variation of $v_{2y}$, on the other hand, exhibits a minor
difference. One orbit comprises of two slightly different half-sinusoids,
rounded and sharp maxima. These correspond to the relatively curved and 
straight parts of the trajectory, respectively. As the dissipation is 
negligible, the antivortex continues in the semi-circular orbit and this is 
reflected in the periodic patterns of $v_{2x}$ and $v_{2y}$. This is also
clearly discernible from Fig. \ref{traj_vd}(a), which is an overlap of 
the snap shots of the vortex and antivortex positions over four orbits. 
We also calculate the velocity profile of the antivortex
analytically using Eq.~(\ref{vel_dip2}). We find that our analytical 
results are in good agreement with the numerical calculations shown in
Fig.~(\ref{vel_comp}).

 Introducing the barrier potential brings out a marked change in $v_{2x}$ 
when the antivortex position has $|y|\leqslant w$. As the antivortex enters
this region $y\in[w,0]$, it experiences a sharp deceleration. 
The velocity then becomes zero at $y = 0$, which is identified based 
on the numerical results. For the $y\approx 0$ domain, the variation of 
$v_{2x}$ is consistent with Eq.~(\ref{vel_x}), where it is shown that 
$v_{2x}\propto y$.  Again, for $y\in[0,-w]$ the antivortex experiences 
acceleration till it leaves the barrier region.
As evident from the Fig. \ref{vel_comp}, the antivortex experiences a large 
change in the velocity when it encounters the barrier and dissipates energy 
through acoustic radiation. This causes an increase in the radius of the 
semi-circular orbit and a corresponding increase 
in time period of the orbit. This is also evident from Fig. \ref{traj_vd}(b), 
which shows both the vortex and antivortex spiral away from the initial 
orbits due to energy dissipation of the vortex dipole through the acoustic 
radiation. As an example, the change in the radius of one of the orbits of the 
antivortex before and after passing through the barrier is indicated with a 
dashed white line in Fig. \ref{traj_vd}(b).

 During the first orbit, unlike  $v_{2x}$, there is little or no effect to 
the $v_{2y}$ in the presence of the barrier. But, there are changes in 
$v_{2y}$ during the later orbits. These are noticeable in 
Fig. \ref{vel_comp}, where there is a periodic variation in 
$v_{2x}$ and $v_{2y}$.

\begin{figure}[h]
 \centering
 \includegraphics[width=8cm]{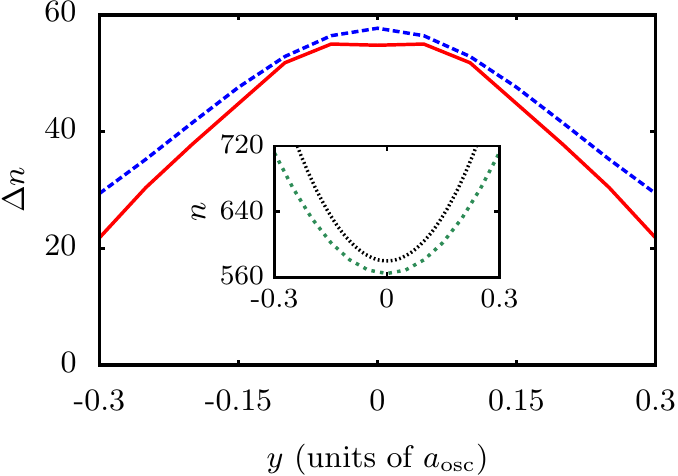}
 \caption{(Color online) The correction to condensate density along
          $y$-direction near the barrier for the quasi 2D BEC. The red-solid 
          line shows the correction to density calculated numerically by 
          subtracting TF solution from the actual GPE solution. The blue-dashed 
          line shows the correction to TF approximation near the repulsive 
          barrier which has been calculated analytically from
          Eq.~(\ref{TF_corr}). Inset: The green-dashed and black-dotted lines 
          are the density profile of the condensate along $y$-direction within 
          the barrier region calculated using numerical computation and 
          analytically from Eq.~(\ref{TF_sol}), respectively. 
          The density is normalized to the total number of atoms $N$ in the 
          condensate.}
 \label{corr_den}
\end{figure}

 In presence of the barrier, as discussed earlier, there is a correction to
the TF density profile within the neighborhood of the barrier potential.
To verify the analytic expression derived earlier, we compare it with the 
results from numerical solution of the GPE. In Fig. \ref{corr_den}, we show 
the plots of the correction to TF-approximation obtained from the numerical 
solution of GPE and estimated using the analytic expression in 
Eq.~(\ref{TF_corr}). It is evident from the figure that the results from the 
analytic expression is in good agreement with the numerical result, at the
center the analytical result is approximately $ 5$\% higher than the numerical 
result.

 It should be emphasized here that the dynamics of a
vortex-antivortex pair also depends on the initial distance between the
constituent vortices as reported in Crasovan 
\emph {et. al}~\cite{crasovan_03}. Depending on the initial position and 
separation, but with the same set of parameters, a range of scenarios are 
possible. In the presence of the barrier, beyond a critical separation 
($\sim 1.6a_{\rm osc}$), the vortex dipole does not cross the barrier and 
move in a closed orbit on one side of the barrier. For lower values of 
vortex-antivortex separation the vortex dipole crosses the barrier and 
dissipates kinetic energy during each crossing. In our present work we 
examine vortex-barrier interaction for a fixed value of vortex-antivortex 
separation, and thus studying the aforementioned dynamics for varying 
vortex-antivortex distance will be investigated in detail in our future 
works.

\begin{figure}[h]
 \centering
 \includegraphics[width=9.0cm]{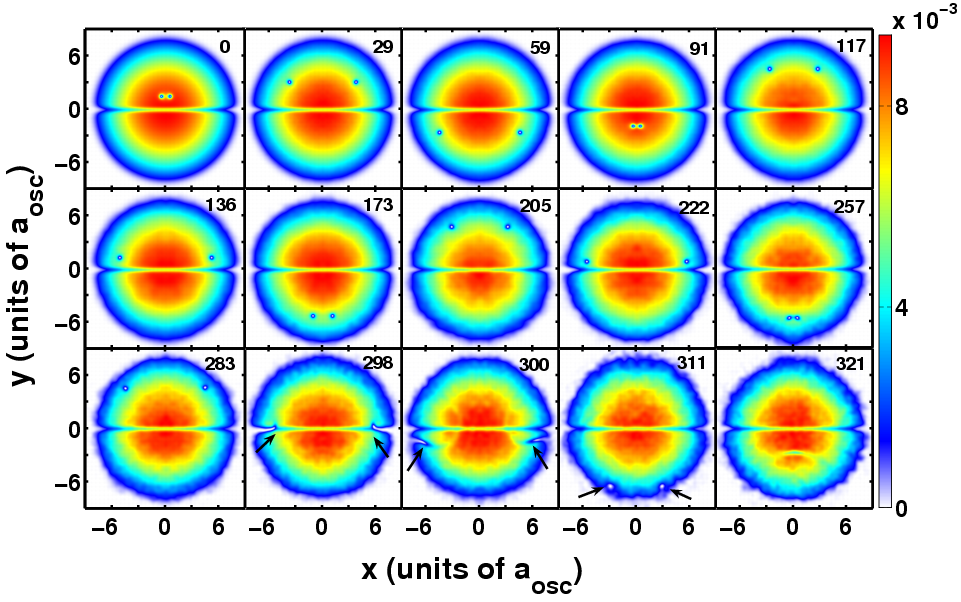}
 \caption{(Color online) Condensate density profiles showing the dynamics
          of a vortex dipole in a harmonically trapped BEC in the presence 
          of the barrier potential.
          The vortex-antivortex pair annihilates at $t = 320$ ms. Density is 
          measured in units of $a_{\rm osc}^{-2}$ and is normalized to unity.
          The vortex-antivortex pair follows the
          trajectories given in the Fig. \ref{traj_vd}(b). In the bottom 
          panels the vortex and antivortex are indicated by the black arrows.
          }
 \label{vortex_anni}
\end{figure}


\subsection{Energetics}
To estimate the energy transferred to the acoustic field by the vortex, 
consider the energy of a vortex located at a distance $b$ from the $z$-axis 
in quasi-2D geometry~\cite{pethick_08}
\begin{equation}
 E \simeq E_0 \left (1 - \frac{b^2}{R_\perp^2} \right )^{3/2},
\end{equation}
where $E_0 = [4\pi\hbar^2 n(0,0) Z/(3m)]\ln (R_\perp/\xi)$ with $Z$ as the 
semi-axis along the $z$-axis and $n(0,0)$ is the density at the origin. 
The above relation is applicable when $Z\gg \xi $ and holds true for the 
parameters we have considered. Based on the above expression, the energy 
radiated by a vortex when the radius of the orbit is increased from $b$ to 
$b+\Delta b$ is 
$ \Delta E \simeq 3(E_0 b/R_\perp ^3)\sqrt{R_\perp^2 - b^2 }\Delta b$. From 
this relation, the energy released by the vortex when it crosses the barrier 
potential in the orbit marked by the dashed line in Fig. \ref{traj_vd}(b) is 
$2.1\times 10^{-3}\hbar\omega_\perp$ and corresponding change in the orbital 
radius $\Delta b$ is $0.5a_{\rm osc}$. It is to be noted that the value of 
$\Delta b$ increases and during the fourth orbit the vortex dipole 
annihilates. For the present case, $V_0= 15\hbar\omega_\perp$, the condensate 
profile at selected time steps are shown in Fig. \ref{vortex_anni}. The 
annihilation of the vortex dipole during the fourth orbital motion is clearly 
discernible in  Fig. \ref{vortex_anni} for the snapshot corresponding to 
321 ms. Considering that the radius $b$ before the annihilation is 
$7.08 a_{\rm osc}$, and vortex and antivortex carry same kinetic energy, the 
energy released during annihilation is $5.21 \times 10^{-3}\hbar\omega_\perp$.

  To study the effect of the barrier on the dynamics of the vortex dipole, 
we further increase $V_0$ to values higher than $15\hbar\omega_\perp$ and 
observe the temporal evolution of the system. At low values, 
$V_0< 25\hbar\omega_\perp$, the vortex dipole crosses the barrier. However, 
the vortex dipole dissipates energy during the crossing and at a later stage, 
the vortex dipole self annihilates. At higher $V_0$, the annihilation of the 
dipole occurs in even less time. But, when $V_0$ is larger than a critical 
value ($25\hbar\omega_\perp$), the vortex dipole does not cross the barrier 
and bounces from the barrier. 

\begin{figure}[h]
 \centering
 \includegraphics[width=8cm]{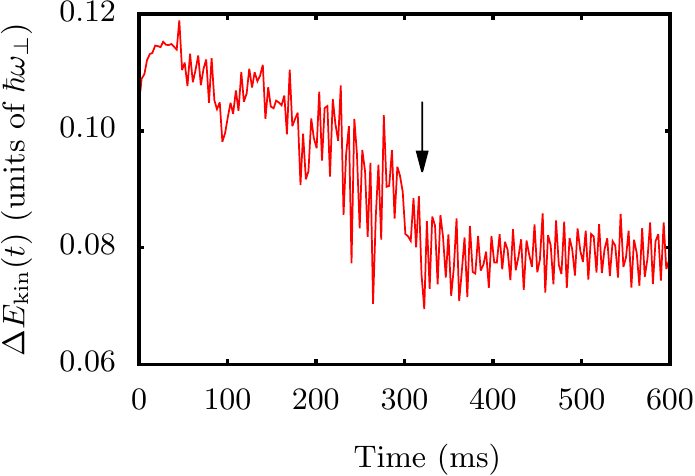}
 \caption{Decay of $\Delta E_{\rm kin}(t)$ due to interaction
           of vortex dipole with barrier in the harmonically trapped
           quasi-2D BEC. The annihilation event at~$320$ ms is indicated by
           an arrow. }
 \label{kin_energy}
\end{figure}

 To analyze the energetics associated with the vortex dipole annihilation,
we examine the contribution from various component to the total energy.
The total energy of the condensate $E_{\rm tot}$, in general, is the sum of 
kinetic energy $E_{\rm kin}$, interaction energy $E_{\rm int}$, quantum 
pressure energy $E_{\rm q}$ and external trapping potential energy 
$E_{\rm ext}$. So, we can write
\begin{equation}
 E_{\rm tot} = E_{\rm kin} + E_{\rm int} + E_{\rm q} + E_{\rm ext},
\end{equation}
where
\begin{subequations}
 \begin{eqnarray}
  E_{\rm kin}(t) & = & \frac{1}{2} \int\big[\sqrt{n(x,y,t)}
                       \mathbf v(x,y,t)\big]^2 dx dy, \\
  E_{\rm int}(t) & = & \int g[n(x,y,t)]^2 dx dy, \\
  E_{\rm q}(t)   & = & \frac{1}{2}\int\big[\nabla\sqrt{n(x,y,t)}
                       \big]^2 dx dy, \\
  E_{\rm ext}(t) & = & \int n(x,y,t) V_{\rm ext} dx dy,
 \end{eqnarray}
\end{subequations}
where the velocity $\mathbf v = \nabla \phi(x,y,t)$. Here, all energies are
in scaled units, as defined earlier. To examine the
kinetic energy of the vortex dipole dissipated, define
\begin{equation}
   \Delta E_{\rm kin}(t) = E_{\rm kin}^{\rm vd}(t) - E_{\rm kin}^0(t),
\end{equation}
where $E_{\rm kin}^{\rm vd}$ and $ E_{\rm kin}^0$ are the kinetic energies 
of the condensate with and without the vortex dipole, respectively. In the 
case of $E_{\rm kin}^{\rm vd}(t)$, it can further be decomposed into a 
component arising from the kinetic energy (KE) of the vortex dipole and a 
part due to the acoustic field. The temporal variation of  
$\Delta E_{\rm kin}(t)$ provides a measure of the
kinetic energy of the vortex dipole transformed into acoustic energy and
is shown in Fig.~\ref{kin_energy}.
After the vortex dipole undergoes annihilation, marked by an arrow in
Fig.~\ref{kin_energy} at $t\sim 320~{\rm ms}$,
a part of the kinetic energy gets
transformed into acoustic radiation energy and the remaining into the 
interaction energy. So, for $t>320~{\rm ms}$ the mean value of 
$\Delta E_{\rm kin}(t)$ after the annihilation is associated with energy 
of the acoustic field in the condensate. In this time domain, the 
$\Delta E_{\rm kin}(t)$ shows rapid variations but are devoid of low-frequency
oscillations associated with the KE of vortex dipole.

\begin{figure}
 \centering
 \includegraphics[width=8cm]{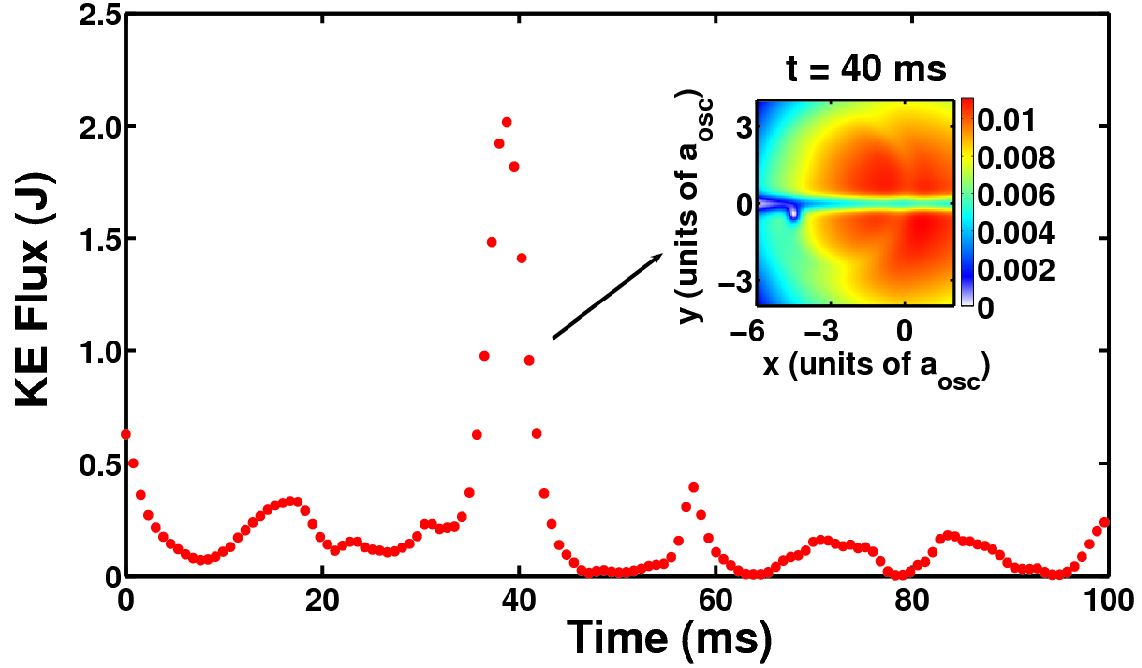}
 \caption{(Color online) The power radiated by a single vortex is shown as a 
          function of time as it passes through a repulsive Gaussian barrier. 
          The inset figure shows the density variation and acoustic wave,
          which is emitted due to vortex-barrier interaction at 
          $t\approx40$ ms. At $t=0$, the vortex is imprinted at 
          $x_0 = 0$,~$y_0 = 3.5$.}           
 \label{pwr_rad}
\end{figure}
 
 We numerically calculate the  kinetic energy lost by the vortex near static 
barrier due to radiation. In order to calculate the KE flux we numerically 
solve Eq.~(\ref{en_flux}) in the far-field approximation. During each 
crossing of the barrier, the vortex emits acoustic radiation 
that significantly perturbs the vortex trajectory. The power radiated due to 
vortex-barrier interaction is shown in Fig. \ref{pwr_rad}, where a large 
decay in energy of the vortex dipole is discernible at~$40$ ms. In addition, 
the vortex also radiates, much lower flux, in the bulk of the condensate 
due to the acceleration arising from the precession. This  is evident from 
the small peaks in the KE flux around~$40$ ms. Due to the radial dependence 
of the acoustic velocity, after emission the acoustic waves acquire 
bow-shaped geometry. At a later time the acoustic waves reflect from the edge 
of the condensate and interact with vortex. In the limit of large $V_0$ the 
vortex reflects back from the barrier and eventually, due to energy loss it 
precess outwards and is lost from the condensate bulk. Apart from radiation 
induced energy dissipation, the broken rotational symmetry of the system 
also affects the life time of the vortex in the condensate~\cite{lundh_00}.


\section{Conclusions}\label{conclusions}

In summary, we studied the effects on a single vortex or a vortex dipole
of a static Gaussian barrier within a harmonically
trapped BEC. We not only examined the dynamics of a vortex dipole, but also
analytically calculated the velocity of the constituent vortices in this 
setting. The analytical results are compared with the numerical solutions and 
we find that these are in good agreement. Unlike in the absence of a barrier, 
we have further demonstrated that the presence of a barrier modifies the 
trajectory of the vortex dipole. Furthermore we have also shown that the 
density anisotropy introduced by the barrier enhances the possibility of 
annihilation events of vortex dipole in atomic BECs through acoustic 
radiation.

 This work opens up the possibility of future investigations on the topic of
vortex-barrier interactions. Among the many interesting prospects that
can be considered are studying the effects of different kinds of barrier on
the vortex dynamics. The vortex-antivortex separation can also be tuned to
study a large number of possible scenarios. Even role of finite temperature
on vortex-barrier interaction is a possible natural extension of the
present work.

\begin{acknowledgments}
We thank S. Gautam and S. Chattopadhyay for very useful discussions. The 
results presented in the paper are based on the computations using the 
3TFLOPs HPC Cluster at Physical Research Laboratory, Ahmedabad, India.
We also thank the anonymous referees for their thorough review and valuable
comments, which contributed to improving the quality of the manuscript.

\end{acknowledgments}

\bibliography{vort_barr}{}
\bibliographystyle{apsrev4-1}

\end{document}